\documentstyle[12pt]{article}
\begin{document}

\voffset -0.4 true cm
\hoffset 1.5 true cm
\topmargin 0.0in
\evensidemargin 0.0in
\oddsidemargin 0.0in
\textheight 8.7in
\textwidth 6.7in
\parskip 10 pt

\def\half{{1 \over 2}}
\newcommand{\ba}{\begin{array}}
\newcommand{\ea}{\end{array}}
\newcommand{\be}{\begin{equation}}
\newcommand{\ee}{\end{equation}}
\newcommand{\bea}{\begin{eqnarray}}
\newcommand{\eea}{\end{eqnarray}}
\newcommand{\beas}{\begin{eqnarray*}}
\newcommand{\eeas}{\end{eqnarray*}}

\begin{titlepage}

\begin{flushright}
IASSNS-HEP-99/18 \\
PUPT-1817 \\
hep-th/9902073 \\
\end{flushright}

\vskip 1.2 true cm

\begin{center}
{\Large \bf Gauge theory origins of supergravity causal structure}
\end{center}

\vskip 0.6 cm

\begin{center}

Daniel Kabat${}^1$ and Gilad Lifschytz${}^2$

\vspace{3mm}

${}^1${\small \sl School of Natural Sciences} \\
{\small \sl Institute for Advanced Study} \\
{\small \sl Princeton, NJ 08540, U.S.A.} \\
\smallskip
{\small \tt kabat@ias.edu}

\vspace{3mm}

${}^2${\small \sl Department of Physics} \\
{\small \sl Joseph Henry Laboratories} \\
{\small \sl Princeton University} \\
{\small \sl Princeton, NJ 08544, U.S.A.} \\
\smallskip
{\small \tt gilad@puhep1.princeton.edu}

\end{center}

\vskip 0.8 cm

\begin{abstract}
\noindent
We discuss the gauge theory mechanisms which are responsible for the
causal structure of the dual supergravity.  For D-brane probes we show
that the light cone structure and Killing horizons of supergravity
emerge dynamically.  They are associated with the appearance of new
light degrees of freedom in the gauge theory, which we explicitly
identify.  This provides a picture of physics at the horizon of a
black hole as seen by a D-brane probe.
\end{abstract}

\end{titlepage}

\section{Introduction}

Given the conjectured dualities between supergravity and gauge theory
\cite{BFSS,ads}, it is natural to ask how the causal structure
of supergravity arises from gauge theory.  In particular we would like
to understand -- from the gauge theory point of view -- why objects in
supergravity must move on trajectories that stay within their future
light-cones.

This question is non-trivial because supergravity generally lives in a
spacetime with more dimensions than its gauge theory dual.  The extra
dimensions of supergravity correspond to the moduli space of vacua of
the gauge theory.  The moduli space carries a Euclidean metric, with
no a priori restriction on how fast an object can travel on it.  So
some dynamical mechanism must be present in the gauge theory to
enforce causality for motion on the moduli space.  The question can be
sharpened by turning on a temperature.  This modifies the causal
structure of the supergravity background, through the appearance of a
black hole with a non-degenerate horizon.  How is this change in
causal structure reflected by the dynamics of finite-temperature SYM?

We first consider supergravity probes, such as dilaton wavepackets in
$AdS_5$, which map to objects that move on the base space of the gauge
theory.  For such probes the supergravity causal structure arises from
kinematics in the SYM: causality in supergravity is related to the
causal structure on the base space by the UV/IR correspondence.  Then
we turn to D-brane probes, made by breaking the gauge group to $SU(N)
\times U(1)$.  These probes move on the moduli space of the gauge
theory.  We show that the restriction on their velocities arises
dynamically, through the appearance of a new light degree of freedom
in the gauge theory: a pair of $W$ particles becomes light as the
velocity of a brane probe approaches the local speed of light.  As a
closely related phenomenon, we also discuss the limiting electric
field on a D-brane probe.  Finally, we turn to black holes, and show
that an isolated $W$ particle becomes massless at the horizon.  This
makes contact with our previous proposal \cite{kl}, where we argued
that a non-extremal horizon is detected by a D-brane probe as the
onset of a tachyon instability\cite{tach}.

To support these claims we use the SYM $\leftrightarrow$ supergravity
correspondence to indirectly calculate the mass of a $W$ in SYM, along
the lines of the Wilson loop computations pioneered in
\cite{ReyYee,MaldaWilson}.  But in the final section of this paper we
perform an explicit diagrammatic calculation of the mass of a $W$.
The diagrammatic calculation is valid at sufficiently high temperature
(at the edge of the region of correspondence with supergravity), and
shows that at least in this special region these claims follow
directly from the SYM itself.

\section{Supergravity probes of $AdS_5$ causality}

We begin with supergravity probes of $AdS_5 \times S^5$.  This is a
simple example in which the light cone structure of the supergravity
arises from kinematics of the gauge theory.  The metric and dilaton are
\begin{eqnarray}
ds^{2} & = & \alpha' \left[\frac{U^{2}}{d_3^{1/2}e}(- dt^2 + dx_{||}^2) +
\frac{d_3^{1/2}e}{U^{2}} \left(dU^2 + U^2 d\Omega_{5}^2\right)\right] \\
e^\phi & = & \frac{g^2_{YM}}{2 \pi} \nonumber
\end{eqnarray}
where $d_3$ is a numerical constant and $e^2 = g^2_{YM} N$.  We need
$e^2$, $N \gg 1$ for the supergravity to be valid.  Consider an object
build out of supergravity fields, at rest on the $S^5$, but with some
velocity in the $U$ direction.  This velocity must satisfy
\begin{equation}
\left\vert \frac{dU}{dt} \right\vert \leq \frac{U^2}{d_3^{1/2}e}
\label{lc1}
\end{equation}
in order that the object not move on a space-like trajectory.

How does such a restriction arise in the SYM?  The dual $SU(N)$ gauge
theory lives on the boundary of $AdS_5$ with coordinates $(t,x_{||})$.
A supergravity excitation at a radial coordinate $U$ corresponds via
holography to an excitation in the SYM with size
\cite{SusskindWitten,BDHM,BalasubKrausLawrTrivedi,PeetPolchinski}
\begin{equation}
\delta x_{||} = \frac{d_3^{1/2}e}{U} \,.
\label{hol1}
\end{equation}
As the supergravity probe moves in the $U$ direction its size changes
in the SYM.  Translating the supergravity condition (\ref{lc1}) into
gauge theory terms using (\ref{hol1}), one finds that causality in
supergravity implies the restriction
\begin{equation}
\left\vert \frac{d\delta x_{||}}{dt}\right\vert < 1\,.
\end{equation}
But this is just the statement that in the gauge theory the size of an
excitation can't change faster than the speed of light.  An equivalent
observation has been made by Susskind \cite{len}, that the time
for a supergravity excitation to propagate directly across $AdS$ space
is equal to the time for a SYM excitation to travel around the
boundary of $AdS$. Other discussions of AdS/CFT causality have been
given in \cite{Das,HorowitzItzhaki,ReyBak}.

So for supergravity probes causal propagation arises as a consequence
of kinematics in the SYM, via the UV/IR relationship (\ref{hol1}).

\section{D-brane probes of causality}

We proceed to study the mechanisms which enforce causality on the
motion of a $p$-brane probe in the near-horizon geometry of an
extremal black $p$-brane.  The dual theory is $p+1$ dimensional SYM in
a certain range of couplings \cite{imsy}, Higgsed to $SU(N) \times
U(1)$.  We will show that causality is enforced dynamically, through
the appearance of new light degrees of freedom in the gauge theory at
a limiting velocity.  We also discuss the closely related phenomenon
of a limiting electric field.

Throughout this section we consider the extremal supergravity background
($e^2 \equiv g^2_{YM} N$)
\bea
ds^2 & = & \alpha'\left[\frac{U^{\frac{7-p}{2}}}{d_{p}^{1/2} e}(-dt^2 +dx_{||}^2)
+\frac{d_{p}^{1/2} e}{U^{\frac{7-p}{2}}}\left(dU^2 + U^2 d\Omega_{8-p}^2\right)\right]
\nonumber \\
e^\phi & = & (2 \pi)^{2-p} g^2_{YM} \left({d_p e^2 \over U^{7-p}}\right)^{3-p \over 4}
\label{ExtremalBackground}
\eea
where
\[
d_p = 2^{7 - 2p} \pi^{9 - 3p \over 2} \Gamma\left({7-p \over 2}\right)
\]
with $B$ vanishing and $N$ units of $(p+2)$-form flux on the
$S^{8-p}$.\footnote{This corresponds to our use of a slightly
non-standard normalization for the SYM coupling, $S_{SYM} = - {1 \over
4 g^2_{YM}} \int {\rm Tr} F^2 + \cdots$.  We thank A.\ Tseytlin for
pointing this out.}

\subsection{Causality and the DBI action}

To establish a framework for our subsequent discussion, we begin by
examining the supergravity effective action for a $p$-brane probe.
This is given by the DBI action \cite{FradkinTseytlin,ACNY,Rob} together
with Chern-Simons couplings \cite{Miao,Mike}.
\begin{equation}
S= - T_{p} \int e^{-\phi} \sqrt{- \det (G+B+F)} + \int C \wedge e^{B+F}
\label{action}
\end{equation}
The effects we will discuss correspond to the restriction that this
action be real, $\det(G+B+F) < 0$.  We would like to
understand how this bound arises from SYM dynamics.

We first consider the case $B + F = 0$, where the restriction simplifies to
$\det G < 0$, the condition for the probe to travel inside its future lightcone.
The action for the probe is
\begin{equation}
S= - \frac{1}{g_{YM}^{2}} \int d^{p+1}x \, \frac{U^{7-p}}{d_{p} e^2}
\left(-1 + \sqrt{1-\frac{d_{p} e^2 }{U^{7-p}}\left(\dot{U}^2
+ U^2 \dot{\Omega}^2\right)}\,\,\right)
\label{action1}
\end{equation}
and the restriction that the brane move on a time-like trajectory 
implies a limit on its velocity
\begin{equation}
\dot{U}^2 + U^2 \dot{\Omega}^2 < \frac{U^{7-p}}{d_{p}e^2}\,.
\label{lc2}
\end{equation}

In the dual SYM one considers an $SU(N+1)$ gauge theory broken to $SU(N)
\times U(1)$ by a Higgs vev $\rho$.  This gives rise to massive W's which
are in the fundamental of $SU(N)$ and charged under the $U(1)$.  We allow
$\rho$ to be time dependent.\footnote{A Higgs vev can't be time dependent
in infinite volume, so we should think of the brane as compactified on a
large torus.}  In the extremal case $\rho$ is identified with
the radial coordinate $U$ of $AdS$, so one has a limitation on how rapidly
a Higgs vev can change in the SYM.  For radial motion
\begin{equation}
\dot{\rho}^2  < {\rho^{7-p} \over d_p e^2} \,.
\label{lc3}
\end{equation}
This is somewhat surprising from the SYM point of view, since the
moduli space does not carry a causal structure, and there is no a
priori kinematical restriction on how fast a probe can move.

We also consider the case of a brane at rest but with a $U(1)$
electric field $F_0{}^i$ present on its worldvolume.  This
is a closely related situation, since an electric field maps
to velocity under T-duality.  From (\ref{action}) one sees that there
is a maximal field strength
\begin{equation}
\sum_i \left(F_0{}^i \right)^2 < 1 \,.
\end{equation}
We seek to understand the SYM origin of these restrictions.

All these restrictions arise because the probe action has a
Born-Infeld term, which becomes singular when the velocity or electric
field gets too large\footnote{The effective action is finite but
non-analytic when the limiting velocity is approached.  But making a
Legendre transform one sees that the Hamiltonian diverges at
criticality: the vacuum energy of the SYM blows up.}.  In string
theory, the Born-Infeld action is the effective action for massless
open strings.  It arises from integrating out massive open string
states at tree level.  In general we expect a singularity of an
effective action to reflect the fact that a new degree of freedom is
becoming light.  Such a phenomenon is well-known in the context of
open string effective actions: in flat space, above a critical
limiting value of the electric field, an open string with
oppositely-charged ends develops a tachyon instability
\cite{burges,nev}.  Intuitively, when the electric field becomes
larger than the string tension the string is unstable and stretches
indefinitely.

The Born-Infeld action is also expected to arise as the effective
action of SYM at large $N$ \cite{bbpt,ct,kp}.  When written in terms
of the 't Hooft coupling $e^2$ note that the supergravity result
(\ref{action1}) is ${\cal O}(N)$.  So it should come from a sum of
planar diagrams with the topology of a disk, where a $W$ runs around
the boundary of the disk and is dressed by $SU(N)$ degrees of freedom
that fill in the interior.  The fact that the resulting effective
action is non-analytic at a critical velocity or electric field
suggests that some degree of freedom which was integrated out becomes
light at the critical field.

It is instructive to regard this non-analyticity as a breakdown of
perturbation theory.  The SYM effective action has a double expansion
in $v^2/\rho^4$ and $g^{2}_{\rm eff}=e^2/\rho^{3-p}$ \cite{bbpt}.  The
diagonal terms in this expansion are expected to re-sum to give the
DBI action (\ref{action1}).  At weak coupling, where $g^2_{\rm eff}
\ll 1$ and a one-loop calculation is reliable, the perturbation series
breaks down when the velocity expansion diverges, at $v^2/\rho^4 \sim
1$.  This is understood as a failure of the Born-Oppenheimer
approximation: $W$ pair creation becomes possible, and gives an
imaginary part to the effective action \cite{bacpor,bachas,dkps}.  On
the other hand, at strong effective coupling $g_{\rm eff}^2 \gg 1$, it
is expected that the perturbation series breaks down when the diagonal
expansion parameter $g_{\rm eff}^2 v^2/\rho^4$ becomes of order one:
this is the point (\ref{lc3}) at which the DBI action is singular.
Note that at strong coupling the diagonal expansion breaks down while
the velocity expansion is still good, since $v^2/\rho^4 \sim
1/g_{eff}^2 \ll 1$.

As we will show in the next section, the state which becomes light as
the probe velocity or electric field is increased is a pair of $W$'s,
neutral under the $SU(N) \times U(1)$.  The mass of an isolated $W$
remains large; it is the interaction between them that makes the pair
light.

\subsection{$W$ masses at strong coupling}

We want to argue that a $W$-pair becomes light as the velocity or
electric field approaches criticality.  This could in principle be
studied directly in the SYM by summing planar diagrams.  In section 5
we will carry out a diagrammatic calculation, valid for sufficiently
high temperatures, but for now we take an indirect approach, and use
the correspondence with supergravity to compute the energy of a pair
at strong coupling.

We do this by adopting methods introduced in \cite{ReyYee,MaldaWilson}
to compute Wilson lines in large-$N$ gauge theory.  From the point of
view of string theory, the worldline of a $W$ is the boundary of a
string worldsheet which has been attached to the probe D-brane.  The
action for the $W$ worldline can be computed by evaluating the
worldsheet action for the corresponding string.  At large $N$ and
large effective coupling we can treat the worldsheet classically,
ignoring fluctuations and string loop corrections.

We consider several instructive examples, which are soluble and serve
to illustrate the behavior we expect in more general situations.
First we examine a $W$ pair in an electric field on a brane at rest.
This provides a guide to the qualitative behavior a string worldsheet
in the backgrounds of interest.  Then we consider a $W$ pair on a
moving brane.  For a particular family of brane trajectories we can
solve for the string worldsheet, and show that the pair becomes
massless exactly at criticality.  Finally we study an isolated $W$ on
a brane moving along an arbitrary trajectory, and show that its mass
remains large even as the velocity is increased.  Although we
explicitly treat the case of an $AdS_5$ background, the same
calculations can be done for any $p$-brane.

\subsubsection{$W$ pair in an electric field}

We seek the classical worldsheet of a fundamental string that starts
and ends on a D3-brane which is located at a fixed position $U=U_f$
and has an electric field $E = F_{0}{}^{1}$ turned on in the $X^1$
direction.  We look for a static solution to the equations of motion
using the ansatz
\beas
X^{0} & = & \tau \qquad \qquad \quad -\infty < \tau < \infty \\
X^{1} & = & L \sigma \qquad \qquad \quad 0 \leq \sigma \leq \pi \\
U & = & U(\sigma)
\label{ansa1}
\eeas
This treatment is essentially identical to \cite{ReyYee,MaldaWilson}; the only
new observation is that by including an electric field we can stabilize the
static solution at a finite value of $U_f$.

The worldsheet action is Nambu-Goto plus a boundary term.\footnote{A
boundary term must be added for the Neumann coordinates when the gauge
field is excited.  But no boundary term is needed in the Dirichlet
directions, for reasons discussed in \cite{Rob}.}  Evaluated on the
ansatz it takes the form
\[
S = - {1 \over 2 \pi} \int d\tau d\sigma \left((U^{'})^2 +\frac{U^4 L^2}{d_3 e^2}\right)^{1/2}
+ {1 \over 2 \pi \alpha'} \int d\tau A_0\big\vert_{\sigma=0}^\pi \,.
\]
As in \cite{ReyYee,MaldaWilson} the action is stationary when
\begin{equation}
\sigma={\pi \over 2} + \frac{d_3^{1/2} e}{U_0 L}\int_1^{U(\sigma)/U_0}
\frac{dy}{y^2\sqrt{y^4-1}}\,.
\label{sol1}
\end{equation}
Here $U_0$ is the minimum value of $U(\sigma)$.  The minimum occurs at
$\sigma = \pi/2$, which gives a relation between $U_0$ and $L$.
\begin{equation}
{\pi \over 2} = \frac{d_3^{1/2} e}{U_{0}L}\int_{1}^{U_{f}/U_{0}}
\frac{dy}{y^2\sqrt{y^4-1}}
\label{con1}
\end{equation}
We must also impose the worldsheet boundary conditions
\be
\partial_{n} X^{\mu}=F^{\mu}{}_{\nu}\partial_{t} X^{\nu}
\label{bndycond}
\ee
where $\partial_n$, $\partial_t$ are the unit normal and tangential
derivatives
\begin{eqnarray}
\partial_{n} & = & \frac{\left(d_3 e^2\right)^{1/4} U_f}{\sqrt{U_{f}^{4}L^2+d_3e^2(U^{'})^2}}
\partial_{\sigma}\\
\partial_{t} & = & \frac{\left(d_3 e^2\right)^{1/4}}{U_{f}}\partial_{\tau}\,.\nonumber
\end{eqnarray}
The boundary conditions are satisfied when $U_0 = E^{1/2} U_f$.  Note
that as we increase the electric field $U_0$ approaches $U_f$, so the
string stays closer to the 3-brane.  But at the same time by
(\ref{con1}) the separation $L\pi$ between the two $W$ particles
decreases.  Note that $L$ goes to zero at the critical field $E = 1$;
this may be an artifact of our static ansatz.

\subsubsection{$W$ pair at criticality}

Consider a brane moving along a trajectory $U = \Phi(X^0)$, and
make the following ansatz for a string worldsheet.
\beas
X^0 & = & \tau \\
X^1 & = & L \sigma \\
U & = & \Phi(\tau)
\eeas
This describes a segment of string lying entirely within the the brane
worldvolume.  The string equations of motion are satisfied provided
the brane trajectory obeys
\[
\ddot{\Phi} - {4 \dot{\Phi}^2 \over \Phi} + {2 \Phi^3 \over d_3 e^2} = 0 \,.
\]
A particularly interesting solution is $\Phi(X^0) = d_3^{1/2} e /
X^0$, which corresponds to a brane moving radially at the local speed
of light.\footnote{This is not physically possible in supergravity, of
course, but we are only using this solution to understand how the same
bound arises in SYM.}

We must also satisfy the boundary conditions (\ref{bndycond}) at the string endpoints.
This can be done by turning on a worldvolume electric field $E =
F_{0}{}^{1}$ which satisfies
\be
E^2 = 1 - {d_3 e^2 \dot{\Phi}^2 \over \Phi^4}\,.
\label{critcond}
\ee
When this condition is obeyed the system is at criticality.  Note that
the length of the string in the $X^1$ direction $L \pi$ remains
arbitrary.  Also note that when the velocity of the brane equals the
local speed of light the electric field necessary to stabilize the
string endpoints vanishes.

Evaluating the action for this configuration we find
\beas
S_{bulk} & = & - {L \over 2} \int d\tau \sqrt{{\Phi^4 \over d_3 e^2} - \dot{\Phi}^2} \\
S_{boundary} & = & {L \over 2} \int d\tau {\Phi^2 \over d_3^{1/2} e} E \,. \\
\eeas
When the boundary conditions (\ref{critcond}) are satisfied the bulk
and boundary actions exactly cancel, showing that a $W$ pair becomes
massless at arbitrary separation in a critical field.

\subsubsection{Mass of an isolated $W$}

As we have seen, a $W$ pair becomes massless when the electric
field and velocity become critical.  This could occur in the gauge
theory for two distinct reasons: either the mass of an individual $W$
particle goes to zero and the interactions are small, or the mass of a
single $W$ remains finite but a large interaction potential cancels
the rest energy of the pair.

To decide between these possibilities we study a single $W$ in
isolation.  This is easiest in the velocity case, with vanishing
electric field.  An isolated $W$ can be obtained from a $W$ pair by
sending the separation $L \rightarrow \infty$ while holding the
velocity fixed.  As can be seen in the example of section 3.2.1, at
large $L$ the worldsheet will tend to adopt a configuration in which
the string stretches from the brane close to the horizon at $U=0$ and
then travels along the horizon before returning back out to the
brane.\footnote{For an extremal background we expect the classical
string worldsheet to change smoothly as $L$ increases, but at finite
temperature the minimal-action worldsheet changes discontinuously at a
finite value of $L$ \cite{ReyTheisenYee,BISY}.}

This means that we can compute the mass of an isolated $W$ by
studying the worldsheet of a string that stretches straight from the
brane to the horizon.  Such a solution is easy to obtain: for a brane
moving on an arbitrary trajectory $U = \Phi(X^0)$ the string
worldsheet is
\beas
X^0(\tau,\sigma) & = & \tau \\
U(\tau,\sigma) & = & {1 \over \pi} \, \sigma \, \Phi(\tau)
\eeas
Evaluating
the Nambu-Goto action for this worldsheet one finds
\[
S = - {1 \over 2 \pi} \int dX^0 \, \Phi(X^0)\,.
\]
The $W$ is at rest in the gauge theory, so we can directly read
off its (time-dependent) mass: $m_W = {1 \over 2 \pi} \Phi$.

Note that $m_W$ is given in terms of the instantaneous Higgs vev by
the usual formula, without any corrections involving the velocity.  In
particular $m_W$ remains finite as the velocity of the brane
approaches the speed of light.  This means that the phenomenon
discussed above, of a $W$ pair at some fixed separation becoming light
at the critical velocity, must be understood as due to a large
interaction energy between the two $W$'s, which cancels their rest
energy in the limit of critical velocity.

Given that $W$ pairs become light as the velocity or electric field is
increased, it is natural to wonder if large amounts of pair production
will take place.  We feel that pair production remains highly
suppressed below criticality, for reasons that are simplest to explain
for electric fields.  Note that although the energy of a $W$ pair
becomes small, the mass of an isolated $W$ remains large, so the pair
is not completely free to separate in the electric field.  Indeed the
static classical solution (\ref{sol1}) suggests that as the electric
field is increased the two $W$ particles are forced to move closer
together.  This reduces their dipole moment, which makes them very
weakly coupled to the electric field.  From the point of view of
string theory, apart from the fact that pair production is suppressed
by $g_s \rightarrow 0$, the $W$ pair corresponds to a neutral string,
which further suppresses pair production \cite{bacpor}.  So we do not
expect much pair production to take place as long as the field is
below critical.

Pair production should set in above the critical field.  For classical
strings this happens abruptly, as soon as the electric field exceeds
the critical value.  But it would be interesting to understand whether
quantum effects can smooth out the transition.  This is an issue of
$1/N$ corrections to the gauge theory.  The classical computations we
have performed are only valid in the strict $N \rightarrow \infty$
limit.  At finite $N$ the $SU(N)$ degrees of freedom can fluctuate,
and it seems reasonable that this corresponds to the light-cone
fluctuations expected in quantum gravity.

\section{D-brane probes of horizons}

It is interesting to extend our discussion of causality to include
finite temperature effects.  On the supergravity side, finite
temperature modifies the causal structure of the spacetime, through
the formation of a black hole with a non-degenerate horizon.  The
presence of this horizon must somehow be reflected in the
finite-temperature dynamics of SYM.

To be more precise, we wish to address the following question.  The
dual gauge theory selects a preferred set of coordinates for
supergravity.  The time coordinate of the SYM corresponds to a
timelike Killing vector of the supergravity background.  In
supergravity this Killing vector can become null, at so-called
Killing horizons.  For example, this always happens at the event
horizon of a stationary black hole \cite{hawkell}, and it also happens
in $AdS_5$ at $U=0$ (which is not a true event horizon).  What is the
signature of a Killing horizon in the gauge theory?

To study this we place a D-brane probe at some Higgs vev $\rho$ in the
gauge theory.  Our proposal is that when the value of the Higgs vev
corresponds to the position of a Killing horizon, the mass of a $W$
goes to zero in the gauge theory.  Note that we are talking about a
single isolated $W$ particle, not the $W$-pair discussed in the
previous section.

This proposal originated in our earlier work \cite{kl}, where we
studied the way in which a black hole absorbs and thermalizes a
D-brane probe.  We argued that the $W$ (mass)$^2$ matrix has a zero
eigenvalue when the Higgs vev takes the value $\rho = \rho_0$ which
corresponds to the position of the horizon.\footnote{At finite $N$ this
is better thought of as the position of the stretched horizon (see
the discussion section).}  For non-extremal black holes we argued
that there is a tachyon instability for $\rho < \rho_0$.  The tachyon
instability gives an imaginary part to the probe effective potential,
which causes absorption, and also makes it possible for the probe to
rapidly thermalize with the black hole.  For extremal black holes a
$W$ goes massless but no tachyon develops, and the absorption
mechanism is more plausibly string pair creation \cite{kl}.

We first consider the case of degenerate horizons (Killing horizons
with zero surface gravity\footnote{but possibly non-zero area}).  For
example, consider the near-horizon geometry of an extremal black
$p$-brane (\ref{ExtremalBackground}).  The timelike Killing vector
$\partial_t$ becomes null at $U=0$.  For a D$p$-brane probe this point indeed
corresponds to the origin of moduli space, where the $W$'s
become massless.  Another example is a D-string probe of an extremal
D1-D5-momentum black hole \cite{dps,MartinecLi}.  Again the timelike Killing
vector becomes null at the point which corresponds to the origin of
moduli space, where new massless particles appear.  These extremal
black holes are supersymmetric, so no tachyon instability can develop,
which means that the thermalization rate for infalling matter is very
slow \cite{kl}.  But this is consistent with the fact that these
degenerate horizons have zero temperature.

Let us now consider the non-degenerate Killing horizons which arise
for non-extremal black branes.  For simplicity we treat the
near-horizon geometry of a ten-dimensional 0-brane black hole,
although the same discussion could be given for any $p$-brane charge
\cite{imsy}.  The string-frame metric and dilaton are
\begin{eqnarray}
ds^{'2} & = & \alpha' \left[-h(U)\frac{U^{7/2}}{d^{1/2}_{0} e}dt^2+h^{-1}(U)
\frac{d^{1/2}_{0}e}{U^{7/2}}dU^2 +
{d^{1/2}_{0}e \over U^{3/2}}d\Omega_{8}^{2}\right] \nonumber \\
e^{\phi} & = & 4\pi^2g_{YM}^{2}\left(\frac{d_{0}e^2}{U^7}\right)^{3/4}
\label{NonExtreme}
\end{eqnarray}
where $h(U)=1-\frac{U_{0}^{7}}{U^{7}}$.  The horizon is located at $U
= U_0$.  

By evaluating the action (\ref{action}) in the background
(\ref{NonExtreme}) we find that the effective action for a 0-brane probe is
\begin{equation}
S = -\frac{1}{g^2_{YM}}\int dt \frac{U^7}{d_{0}e^2}
\left(-1 + \sqrt{h(U)-\frac{\dot{U}^{2}d_{0}e^2}{h(U)U^7}}\,\,\right)\,.
\label{NonExtAction}
\end{equation}
Thus there is a limit on the radial velocity of the probe
\[
{dU \over dt} < {h(U) U^{7/2} \over d_0^{1/2} e}\,.
\]

The simple identification between the $U$ coordinate of supergravity
and the Higgs vev $\rho$ of the gauge theory no longer holds when the
black hole has finite temperature.  Rather, to get the
kinetic terms to agree, one must set
\cite{maldaprobe,ty}
\begin{equation}
\frac{d \rho}{\rho}= \frac{d U}{U\sqrt{h}}\,.
\end{equation}
This gives the relation
\begin{equation}
\rho^{7/2} = \half \left(U^{7/2} + \sqrt{U^7 - U_0^7}\right)\,.
\label{rho}
\end{equation}
At extremality this reduces to the expected result $\rho = U$, and we
see that an extremal horizon indeed maps to the origin of moduli
space.  But away from extremality the horizon corresponds to a
non-zero Higgs vev $\rho_0 = \rho(U_0) \sim U_0$.  Note that the
interior of the horizon $U < U_0$ in supergravity corresponds to
imaginary Higgs vevs in the gauge theory.

The restriction on velocity translates by (\ref{rho}) 
into a restriction on the rate of change of the Higgs vev
\begin{equation}
\frac{d \rho}{dt} < \frac{\rho^{7/2}}{d^{1/2}_{0}e}
\frac{(1-\frac{\rho_{0}^{7}}{\rho^{7}})}{(1+\frac{\rho_{0}^{7}}
{\rho^{7}})^{2/7}}\,.
\end{equation}
Away from the horizon the physics of this bound is essentially the same as in
the previous section, with a W-pair becoming massless due to
interactions when the velocity approaches its limiting value.

But note that even for zero velocity the effective action
(\ref{NonExtAction}) is singular at the horizon, where $h(U)$
vanishes.\footnote{The effective action as a function of $\rho$ looks
analytic, but $\rho(U)$ is non-analytic at the horizon.}  This
non-analyticity signals the breakdown of the Born-Oppenheimer
approximation.  But unlike the situation in section 3, the breakdown
is now due to an isolated $W$ becoming massless.  To show this, we
proceed as in section 3.2.3, and calculate the mass of an isolated $W$
from the worldsheet action for an elementary string stretched between
the probe and the horizon.  This is identical to the calculations
performed in \cite{ReyTheisenYee,BISY} in the context of
finite-temperature Wilson lines.  For a brane located at $U =
\Phi(X^0)$ this gives the $W$ mass $m_W = \left(\Phi -
U_0\right)/2\pi$.  We see that a single $W$ indeed becomes massless at
the horizon.  Extrapolation to $\Phi < U_0$ would give an imaginary
mass, since $U$ becomes a timelike coordinate inside the horizon.

\section{Causality in M(atrix) theory}

For non-conformal $p$-branes the supergravity background
(\ref{ExtremalBackground}) is only valid for a limited range of the
radial coordinate \cite{imsy}.  In particular, for a system of
D0-branes, the description in terms of ten-dimensional supergravity
breaks down at short distances, and must be replaced by M-theory.  We
now briefly discuss causality in this regime.

A minor point is that M(atrix) theory \cite{BFSS} has a decoupled $U(1)$ sector
describing center of mass motion.  In the c.m. sector the theory is
free and there is no limit on how rapidly a scalar field can change.
But this is consistent with the fact that M(atrix) theory is a
light-front description of M-theory.  In light front coordinates the
requirement of causality\footnote{An overdot denotes $\partial_\tau$,
where $\tau = \half x^+ = \half\left(t + x^{11}\right)$ as in
\cite{bbpt}.}
\[
\dot{x}_\perp^2 < 2 \dot{x}^-
\]
puts no restriction on the transverse c.m. velocity: for any
$\dot{x}_\perp$ this inequality can be satisfied simply by making
$\dot{x}^-$ sufficiently large.

A restriction does arise, however, on the transverse relative
velocity.  This is because relative motion is governed by the smeared
Aichelburg-Sexl metric in eleven dimensions \cite{bbpt}
\be
ds^2 = - dx^+ dx^- + dx_\perp^2 + {15 N \over 2 R^2 M^9 r^7} \left(dx^-\right)^2 \,.
\ee
Causality then requires
\[
\dot{x}_\perp^2 < 2 \dot{x}^- - {15 N \over 2 R^2 M^9 r^7} \left(\dot{x}^-\right)^2\,.
\]
This can only be satisfied for some $\dot{x}^-$ provided that the transverse relative
velocity is bounded,
\[
\dot{x}_\perp^2 < {2 R^2 M^9 r^7 \over 15 N}
\]
or equivalently
\be
\dot{U}^2 +U^2 \dot{\Omega}^2 < {U^7 \over 240 \pi^5 e^2}
\label{MatrixBound}
\ee
where we have used $R M^3 = 1/2\pi\alpha'$ and $R = (2 \pi \alpha')^2
g^2_{YM}$.  This agrees precisely with the bound (\ref{lc2}) extracted
from the ten dimensional supergravity background.

This agreement isn't surprising, since the classical supergravity
backgrounds are so closely related.  But it does suggest that
$W$-pairs become massless in the gauge theory at the critical velocity
(\ref{MatrixBound}), even in the M(atrix) theory regime.

\section{Direct calculation of the $W$ mass}

In this section we discuss the direct evaluation of the $W$ mass for a
system of D0-branes at a particular temperature, and show that it
agrees with the supergravity result for a black hole of the same
temperature.  The supergravity is only valid up to black hole
temperatures of order $T_{c} \sim e^{2/3}$, because at this
temperature the curvature at the horizon becomes of order one in
string units \cite{imsy}. Above $T_c$ the gauge theory is weakly
coupled. Therefore at $T \sim T_{c}$ we should be able to
qualitatively match perturbative SYM calculations onto supergravity
calculations.

We wish to compute the thermal partition function of 0+1 dimensional
SYM, with Euclidean action \cite{ClaudsonHalpern,BFSS} ($e^2 =
g^2_{\rm YM} N$)
\bea
S_{SYM} & = & {1 \over g^2_{YM}} \int_0^\beta d\tau \,\, {\rm Tr} \biggl\lbrace
\half \partial_\tau X^i \partial_\tau X^i - {1 \over 4} [X^i,X^j] [X^i,X^j]
\nonumber \\
& & \qquad \qquad \qquad \quad
+ \Psi_a \partial_\tau \Psi_a + \Psi_a \gamma^i_{ab}
[X^i,\Psi_b] \biggr\rbrace \,.
\label{SSYM}
\eea
We work in the high temperature regime $T > T_c$.  At high
temperatures the gauge theory is weakly coupled, so the naive
expectation is that the partition function can be obtained from the
tree-level spectrum of massless bosons and fermions.  This logic fails
for the bosons, because higher loop corrections, although suppressed
by powers of the coupling, are in fact infrared divergent due to the
bosonic zero modes.\footnote{At finite temperature the fermions are
antiperiodic and do not have zero modes.}

These infrared divergences are cured through the generation of a
thermal mass for the bosons.  To obtain the mass gap we must re-sum
part of the perturbation series.  This can be done as follows.  As we
argue below, at high temperatures the fermions make a negligible
contribution to the mass gap.  So we temporarily ignore them, and
approximate the bosonic sector of the SYM action (\ref{SSYM}) by the
Gaussian action
\be
S_0 = \sum_{l=-\infty}^{\infty}\frac{1}{2\sigma_{l}^{2}}{\rm Tr}\left(X^{i}_{l}X^{i}_{-l}\right)\,.
\label{essnot}
\ee
Here $l$ labels Fourier modes in the Euclidean time direction.  To fix the
parameters $\sigma^2_l$ we make a large-$N$ variational approximation
\cite{el}.  This gives a gap equation for the effective propagator
\begin{equation}
\frac{1}{\sigma_{l}^{2}}={1 \over g^2_{\rm YM}} \left(\frac{2 \pi l}{\beta}
\right)^2 + \frac{16N}{g^2_{\rm YM} \beta}
\sum_{m} \sigma_{m}^{2}.
\label{bb}
\end{equation}
By scaling this equation one can show that the dimensionless effective
propagator at large $N$, namely $\widetilde{\sigma}^2_l = N \sigma_l^2
/ e^{2/3}$, only depends on the dimensionless effective temperature
$\widetilde{T} = T/e^{2/3}$.

The second term on the right hand side of the gap equation is a
thermal mass for the bose fields.  In fact this gap equation takes into
account all leading high-temperature corrections to the bose
propagator.  In writing the gap equation we have ignored corrections
to the propagator involving a single fermion loop, as well as
multiloop corrections.  But the fermion loop is negligible at high
temperatures (it vanishes in the high temperature limit, due to the
antiperiodic boundary conditions for the fermions).  And multiloop
corrections to the propagator are small at high temperatures, where
the theory is weakly coupled.\footnote{Once the thermal mass has been
taken into account multiloop corrections truly are suppressed at weak
coupling.}

To solve the gap equation we introduce the `size' of the thermal
state, defined by the range of eigenvalues of the matrices $X^i$.
\beas
R_{\rm rms}^{2} & \equiv & {1 \over N} <{\rm Tr}\, \left(X^i(\tau)\right)^2 >
\qquad \quad \hbox{\rm (no sum on $i$)} \\
\noalign{\vskip 2 mm}
& = & \frac{N}{\beta} \sum_l \sigma^{2}_{l}
\eeas
The gap equation gives a consistency condition which determines $R_{\rm rms}$.
\begin{equation}
R_{\rm rms}^3 = {e^2 \over 8 \tanh \left(2 \beta R_{\rm rms}\right)}
\label{bb1}
\end{equation}
In this approximation the SYM consists of $9 N^2$ bosonic harmonic oscillators
with frequency $4 R_{\rm rms}$, plus $8 N^2$ decoupled massless
fermions, so the SYM free energy at high temperatures
is given by\footnote{There are also ghosts
associated with the choice of axial gauge $A_0 = 0$, but their contribution
to the partition function is just $\det' \partial_\tau = 1$.}
\begin{equation}
\beta F_{SYM} \approx 9 N^2 \log [2\sinh(2 \beta R_{\rm rms})] - 8 N^2 \log 2.
\label{GaussianBF}
\end{equation}
Thus at high temperatures ($T \gg e^{2/3}$) we have the approximate
size, energy and entropy
\begin{eqnarray}
\label{ther}
R_{\rm rms} & \approx & \frac{1}{2} e^{1/2} T^{1/4} \nonumber \\
E_{SYM} & \approx & {27 \over 4} N^2 T \\
S_{SYM} & \approx & N^2 \left({\rm const.} + {27 \over 4} \log (T/e^{2/3})
\right) \nonumber
\end{eqnarray}
These are the standard results for perturbative SYM in this temperature
regime.

The correct results in the low temperature regime $T \ll e^{2/3}$ can be extracted
from the dual supergravity background (\ref{NonExtreme}).  One finds that
the horizon radius, energy and entropy are given by \cite{imsy}
\begin{eqnarray}
U_{0} & \sim & e^{2/3}\left(\frac{T}{e^{2/3}}\right)^{2/5} \nonumber \\
E_{SUGRA} & \sim & N^2 e^{2/3}\left(\frac{T}{e^{2/3}}\right)^{14/5} \\
S_{SUGRA} & \sim & N^2 \left({T \over e^{2/3}}\right)^{9/5} \nonumber
\end{eqnarray}
We see that the supergravity results for the energy and entropy agree
with the perturbative SYM results (\ref{ther}) at $T \sim e^{2/3}$.
Moreover, the Higgs vev $\rho_0$ at which a
probe 0-brane develops a tachyon instability in the Gaussian ensemble
(\ref{essnot}) can be computed, and one finds that \cite{kl}
\begin{equation}
\rho_0 \left(T=e^{2/3}\right) \sim e^{2/3}\,.
\end{equation}
This agrees with the supergravity result that $U_{0}\vert_{T=e^{2/3}} \sim
e^{2/3}$.  Thus on the edge of validity of the supergravity
description one can apply a Gaussian approximation to the SYM, and one
finds that the horizon of the black hole indeed coincides with the
onset of a tachyon instability for a 0-brane probe.

Directly analyzing the SYM at low temperatures is a difficult problem.
But the bosonic sector of the gauge theory is easier to study: the
Gaussian approximation (\ref{essnot}), (\ref{bb}) is a good
description of bosonic large-$N$ Yang-Mills quantum mechanics at any
temperature.\footnote{One can calculate the leading corrections to the
Gaussian approximation for the bosonic partition function and show
that they are small.}  Thus for the bosonic theory we have
\[
\beta F_{\rm bose} = 9 N^2 \log [2\sinh(2 \beta R_{\rm rms})]
\]
and in the low temperature regime ($T \ll e^{2/3}$)
\begin{eqnarray}
\label{therb}
R_{\rm rms} & \approx &  \half e^{2/3} \nonumber \\
E_{\rm bose} & \approx & 9 N^2 e^{2/3} \\
S_{\rm bose} & \sim & N^2 \gamma(T) \nonumber
\end{eqnarray}
where $\gamma(0) = 0$ and $\gamma(e^{2/3}) = 1$.  The energy is just
the ground state energy of the bosonic harmonic oscillators.  Notice
that even at zero temperature the ground state fluctuations $\sim
R_{\rm rms}$ in the bosonic theory are very large.  In fact they
extend out to the edge of the region $\sim e^{2/3}$ in which (in the
supersymmetric case) supergravity is valid.

At temperatures below $e^{2/3}$ the fermions in the SYM are crucial
for canceling the enormous ground state energy of the bosons which
appears in (\ref{therb}).  While inclusion of the fermions will lower
the ground state energy it cannot significantly alter $<{\rm Tr}\,
X^2>$, which continues to correspond to the size of the entire region
in which supergravity is valid.  So generically all the interesting
supergravity physics, and in particular the horizon radius $U_0$, lies
deep inside the region of the bosonic ground state fluctuations
\cite{len,polun}.

\section{Conclusions and Discussion}

We have analyzed the emergence of causal structure in supergravity in
view of the SYM $\leftrightarrow$ supergravity correspondence.  We
have shown that for D-brane probes the supergravity restriction, that
a probe travel slower than the local speed of light, arises dynamically
in the gauge theory, through the appearance of a light composite state
of two $W$'s.  The same phenomenon is responsible for the limitation
on electric field.  We have also confirmed that a Killing horizon in
supergravity is associated with a $W$ becoming massless, and that
with non-zero surface gravity the $W$ may further become tachyonic.
It is amusing to see that whenever a D-brane probe tries to move on a
lightlike path, a massless mode appears in the SYM.  This gives a
coherent and simple picture of the origin of supergravity causal
structure for D-brane probes.

We would like to comment on the picture of black hole physics that
seems to arise.  It has been argued \cite{thof,stu} that there should
be a unitary description of black hole formation and evaporation from
the point of view of an external observer.  For this to happen the
semiclassical approximation to quantum gravity (i.e. matter fields
propagating on a fixed background) should break down near the black
hole horizon even though the curvature there is small. The black hole
should have a ``stretched horizon,'' at a radius slightly larger than
the radius of the event horizon.  Hawking radiation emerges from the
stretched horizon which is endowed with the thermal properties of the
black hole.  It can be shown \cite{break} that the semiclassical
approximation breaks down if one uses spacetime foliations associated
with outside observers.  In this scheme the stretched horizon arises
naturally as the locus of points where the semiclassical approximation
breaks down. It was also argued that the description of physics as
seen by observers falling into the black hole is complementary to the
description of the black hole as seen by the external observers.

In the supergravity $\leftrightarrow$ SYM correspondence one
identifies the SYM time coordinate with the timelike Killing
coordinate outside the black hole horizon. This means that the SYM
description of a black hole state corresponds to an outside observer's
description of black hole physics.  In classical supergravity,
corresponding to leading $1/N$ effects in the SYM, the stretched
horizon and event horizon cannot be distinguished.  This is the limit
we have considered in this paper.  Subleading corrections, however, do
distinguish them.  At finite $N$ the thermodynamic limit is replaced
by statistical mechanics, and fluctuations can occur.  In particular
the value of the tachyon instability radius will fluctuate.  It seems
natural to us to associate the outer limit of the tachyon instability
with the stretched horizon.  This is why a probe will thermalize at
the stretched horizon, where the semiclassical approximation breaks
down.  As the gauge theory develops a massless mode at the stretched
horizon, the IR/UV correspondence suggests that physics there is
influenced by quantum gravity effects.

Below the Higgs vev corresponding to the stretched horizon, the gauge
theory has no space time interpretation.  Instead the probe has
entered a region where the non-Abelian degrees of freedom control the
physics.  Thus for the outside observer the region behind the
stretched horizon has no spacetime interpretation, unlike the
situation for an infalling observer.

\vskip 1.0 cm
\centerline{\bf Acknowledgements}

We are grateful to Constantin Bachas, Misha Fogler, Michael Gutperle, Roman
Jackiw, Anton Kapustin, Igor Klebanov, Miao Li, Samir Mathur, Vipul
Periwal, Joe Polchinski, Steve Shenker, Lenny Susskind, Sandip Trivedi and
Arkady Tseytlin for valuable discussions.  The work of DK is supported by
the DOE under contract DE-FG02-90ER40542 and by the generosity of Martin
and Helen Chooljian. The work of GL is supported by the NSF under grant
PHY-98-02484.

\end{document}